\documentclass[aps,superscriptaddress,showpacs,floatfix,amsmath,amssymb,twocolumn]{revtex4}

\usepackage{epsfig}
\usepackage{dcolumn}
\usepackage{bm}
\usepackage{color}

\begin{document}
\title{Shell Evolutions and Nuclear Forces }
%\subtitle{Talk given at the 25$^{th}$ International Nuclear Physics Conference (INPC), Firenze, Italy, 2-7 June 2013\\}

\author{O.~Sorlin}
\affiliation{Grand Acc\'el\'erateur National d'Ions Lourds (GANIL),
CEA/DSM - CNRS/IN2P3, BP 55027, F-14076 Caen Cedex 5,
France}

\begin{abstract}%
During the last 30 years, and more specifically during the last 10 years, many experiments have been carried out worldwide using different techniques to study the shell evolution of nuclei far from stability. What seemed not conceivable some decades ago became rather common: all known magic numbers that are present in the valley of stability disappear far from stability and are replaced by new ones at the drip line. By gathering selected experimental results, beautifully consistent pictures emerge, that very likely take root in the properties of the nuclear forces.The present manuscript describes some of these discoveries and proposes an intuitive understanding of these shell evolutions derived from observations. Extrapolations to yet unstudied regions, as where the explosive r-process nucleosynthesis occurs, are proposed. Some remaining challenges and puzzling questions are also addressed.

\end{abstract}
\maketitle

$\it{Foreword}$: The present manuscript is a summary of a plenary talk given at the 25$^{th}$ International Nuclear Physics Conference (INPC), Firenze, Italy, 2-7 June 2013. Some new remarkable experimental results released since then have been implemented here. 

\section{Introduction}
\label{intro}

As early as 1934, W.~Elsasser noticed the existence of "special numbers" of neutrons and protons that confer to the
corresponding nuclei a particularly stable configuration~\cite{Elsa34}. In analogy with atomic electrons, he correlated these numbers with closed shells in a model of non-interacting nucleons occupying energy levels generated by a potential well. It took 15 more years until Mayer, Haxel, Suess and Jensen~\cite{Goep49,Haxe49} showed that this nuclear potential could be modeled by a one-body Harmonic Oscillator (HO), an L$^2$ term and a spin-orbit (SO) potential to  give rise to the HO shell gaps at 8, 20, 40 and SO shell gaps at 28, 50, 82 and 126. In 1954, Elliott wrote that this $\textit{one-body}$ spin-orbit "term may be a caricature of a more  complicated force" as it must be changed when particular shells are filled~\cite{Eliot54}. To give an example, using yet available experimental data, the N=14, 28 and 50 SO shell gaps increase by the \textit{same} value of about 2.7 MeV when the neutron orbits d$_{5/2}$, f$_{7/2}$ and g$_{9/2}$ are filled, respectively(see Fig. 10 in~\cite{Sorlin}). This points to a universal mechanism that cannot be accounted for by one-body potentials for which a single parametrization cannot lead to such modifications in a given isotopic chain. Indeed recent theoretical achievements point out the role of two and even three-body interactions to account for these behaviors~\cite{Holt,Sieja}. In 1960, Talmi and Unna~\cite{Talm60} noticed that nuclear forces can modify the ordering of the shells, stressing that "the order of the filling of the neutron shells  may depend on the proton configuration". These combined works of Elliott and Talmi and Unna introduced the possibility of shell evolutions due to specific two-body nuclear forces.

It took however about 20 more years until the combined works of Thibault~\cite{Thib75} , Hubert~\cite{Hube78}, D\'etraz~\cite{Detr79} and Guillemaud-Mueller~\cite{Guil84} on atomic masses, nuclear radii and nuclear spectra demonstrated the first disappearance of a  magic shell (N=20) and the onset of collectivity far from stability. On the theoretical point of view the works of Campi, Poves and Warburton~\cite{Camp75,Pove87,Warb90} described this onset of collectivity as due to the combination of a shell gap reduction and the excitations of particles from the normally occupied orbital to the first orbital of the upper shell, often referred to as \emph{intruder orbit}.  Some nuclei, in which the ordering of the intruder and normal configurations is inverted belong to the so-called "Island of Inversion"~\cite{Warb90}. 
 These pioneering discoveries in the N=20 region opened a new branch of experimental and theoretical researches related to the structure of nuclei far from stability. This triggered the development  of radioactive ion beam facilities and of highly sensitive detector arrays, that in turn bring new discoveries in other regions of the chart of nuclides. This territorial exploration of the chart of nuclides goes in concert with the inspection of various facets of the nuclear force as different combinations of protons and neutron orbits are filled, from the proton to the neutron drip lines. 
 
 %On the theoretical side, shell model or mean field models based on effective interactions were used to compare to experimental results, aiming at a better understanding and predictability of the evolution of nuclear structure and correlations towards the drip lines. During the last decade two additional major theoretical breakthrough occurred. First, ab-initio models using realistic nucleon-nucleon interactions start to model successfully the spectroscopy of atomic nuclei around closed shells. Second the modeling of open quantum systems such as unbound resonances is being developed [CHECK MAREK]. 
Combining all experimental discoveries, some remarkable and general shell evolutions can be observed in the chart of nuclides as the one evoked above about the N=14, 28 and 50 shell gaps. In the present contribution, the striking analogy between the behavior of the HO gaps N=8, 20 and 40 from the valley of stability to more neutron-rich regions will be presented in Sect.~\ref{HO}. The onset of deformation through intruder configurations at N=20 will be depicted in Sect.~\ref{island} using two recent experimental studies on the $^{34}$Si and $^{32}$Mg isotones. Sect.~\ref{N17} proposes to study the evolution of the neutron single particle energies of the neutron d$_{3/2}$, f$_{7/2}$ and p$_{3/2}$ orbitals between which the N=20 and N=28 gaps are formed. The underlying nuclear forces leading to the disappearance of the N=20 shell gap and the swapping between the f$_{7/2}$ and 
p$_{3/2}$ orbit will be described. Sect.~\ref{hierarchy} shows that a hierarchy in the nuclear forces is responsible for these drastic shell evolutions. A generalization of the hierarchy of nuclear forces  mechanism to other regions of the chart of nuclides (around $^{60}$Ca, below $^{78}$Ni and below $^{132}$Sn where the r-process nucleosynthesis occurs) is  proposed. Sect.\ref{drip} addresses the role of continuum in changing 'effective' nuclear forces in close-to-drip-line nuclei.

\section{Remarkable properties of the HO magic numbers}
\label{HO}

\begin{figure}
\centering
\includegraphics[width=8.3cm]{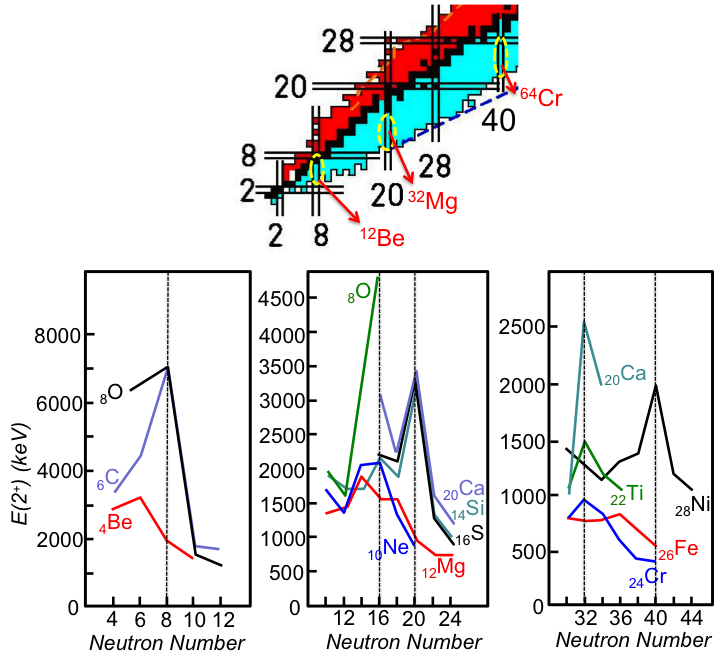}
\caption{Top: The parts of the chart of the nuclides in which the same mechanisms leading to the islands of inversion occurs are indicated by yellow dashed ellipsoids. Bottom: Evolution of the 2$^+_1$ energies in the N=8 (Be, C, O), 20 (O, Ne, Mg, Si, S and Ca) and 40 (Ni, Fe, Cr, Ti, Ca) regions. The rise in 2$^+_1$ energies at N=8, 20 and 40 no longer exists for neutron-rich nuclei. Rather rises in 2$^+_1$ energies appear at N=16 and N=32.}
\label{E2plus}       
\end{figure}

There are several means to test the existence of neutron shell gaps, based on (i) the search of sudden drops in neutron separation energy between adjacent nuclei, (ii) the study of evolution of neutron single particle energies, (iii) the discovery of rises in 2$^+_1$ energy and (iv) reduction in the B(E2; 2$^+_1 \rightarrow $0$^+_1$) values in a given isotopic chain, to quote a few. As shown in Fig.~\ref{E2plus}, the increase of the 2$^+$ energies at the neutron numbers N=8,20 and 40 reveals the presence of large gaps between the last occupied neutron orbit and the valence one. By looking at the same systematics for neutron-rich nuclei in the $_4$Be,  $_{10}$Ne and $_{12}$Mg, $_{26}$Fe and $_{24}$Cr chains there is no more rise of the 2$^+_1$ energies at any of the N=8, 20 and 40 neutron numbers. This partly arises from the fact that the neutron shell gaps have been reduced there, but also from the fact the proton (sub-)shells are no longer fully filled, bringing additional correlations provided by this open proton shell.  From the similarity of these observations at N=8, 20 and 40, we can anticipate that the same process is occurring in all these regions.  Note for instance that, since the pioneering works of Refs~\cite{N40p}, many recent experimental studies have been carried out to study the N=40 sub-shell closure worldwide this last decade~\cite{N40}. We shall demonstrate in Section.~\ref{hierarchy} that the quantum characteristics of the orbits in which protons are added do matter for the creation of shell gaps, by specific proton-neutron interaction,  from the neutron drip line to the valley of stability. Remarkably, far from stability, before these proton-neutron forces act, other local magic numbers at N=16 and N=32 exist, as proposed in Refs.~\cite{Oza00,Hoff,Tshoo,Huck85,Wien}. In the next Section we study how abrupt is the transition to deformation in the N=20 isotones by looking at the systematics of 0$^+_2$ energies.
%The nuclei $^{12}$Be$_{8}$, $^{32}$Mg$_{20}$ and $^{64}$Cr$_{40}$ are deformed, a feature that is confirmed by their large %B(E2) values. 
\section{The Island of Inversion}
\label{island}
The 0$^+_1$ ground state of even-even magic nuclei  can be described, in a first approximation, with all nucleons paired in the occupied orbitals, the valence orbit higher up in energy being unoccupied. It is known however that, for instance, correlations of pairing origin can slightly dilute the occupancy of pairs to the upper shell. The main configuration of the 0$^+_2$ states  corresponds to the promotion of a pair of nucleons to the upper shell (labelled as a 2p2h configuration). The energy of this state is twice the energy to promote a nucleon across the shell gap minus the correlation energy gained by the particles and holes. The constancy of the energy of the  0$^+_2$  at N=20 in the $_{20}$Ca, $_{18}$Ar and $_{16}$S nuclei (E(0$^+_2)\simeq$3.5MeV) indicates that the size of the neutron shell gap (formed between the d$_{3/2}$ and f$_{7/2}$ or p$_{3/2}$ orbits) and the amount of correlations are unchanged between Z=20 and Z=16. With a reduction of the shell gap and the increase of correlations due to the open proton orbit d$_{5/2}$, an inversion of the 0$^+_1$ and 0$^+_2$ states should be occurring as predicted more than 30 years ago. 
\begin{figure}
\centering
\includegraphics[width=8.3cm]{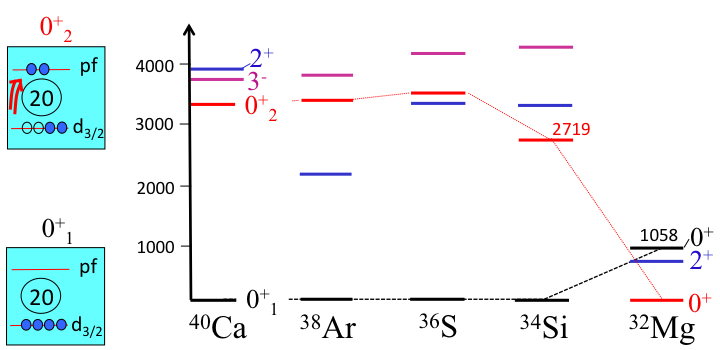}
\caption{ Evolution of the 2$^+_1$ and 0$^+_2$ energies in the N=20 isotones. An abrupt inversion between the 0p0h and 2p2h configurations occurs between the $^{34}$Si and $^{32}$Mg nuclei.}
\label{0plus}       
\end{figure}
To ascertain this mechanism, the energies and configurations of the 0$^+_1$ and 0$^+_2$ in the $^{34}$Si$_{20}$ and $^{32}$Mg$_{20}$, between which the transition happens, should be determined.  The study of these 'missing links' was achieved only these last three years. The $^{30}$Mg(t,p)$^{32}$Mg reaction was used at the Rex-Isolde facility to identify the 0$^+_2$ in $^{32}$Mg at 1058 keV~\cite{Wimm}. Protons were detected in a highly segmented Si detector array (T-ReX), while the $\gamma$-rays following the 0$^+_2 \rightarrow $2$^+_1 \rightarrow $ 0$^+_1$ transitions were observed in the Miniball detector array. The 0$^+_2$ being produced with a similar cross section than the ground state  0$^+_1$, one can think that being close in energy, the two 0$^+$ interact and become largely mixed. The 0$^+_2$ in $^{34}$Si was produced at the GANIL facility through the $\beta$-decay of a newly discovered isomeric state J=1$^+$ of 26(1) ms in the mother nucleus $^{34}$Al, whose neutron configuration 2p1h favors the feeding of a 2p2h configuration in $^{34}$Si through the conversion of a neutron d$_{3/2}$ into a proton d$_{5/2}$~\cite{Rota}. The selective decay of this 1$^+$ isomer though a Gamow Teller $\beta^-$-decay ($\Delta$ J=0 $\pm$ 1) was the key of success for populating the 0$^+_2$ state at 2.72 MeV that decays by pair creation with a half-life of 19.4(7) ns to the  0$^+_1$ state.  The 2$^+_1$ state at 3327 keV was also fed in this $\beta^-$ decay. Information on its competing decay branches to the 0$^+_2$ and 0$^+_1$ was used to infer that the mixing between the two 0$^+$ states is of the order of 20\%~\cite{Rota}. The 0$^+_1$ ground state has 80\% 0p0h and 20\% 2p2h configuration while that of the 0$^+_2$ at 2.72 MeV is reversed. As shown in Fig.~\ref{0plus} an abrupt transition from spherical to deformed configurations is observed between $^{34}$Si and $^{32}$Mg, both from the trend in 2$^+_1$ and 
0$^+_2$ energies. 

\section{Nuclear structure in the N=17 isotones}
\label{N17}
Concomitant information from the S$_{2n}$, 2$^+_1$, 0$^+_2$ energies and B(E2) systematics suggest an erosion of the N=20 shell gap. But these global properties do not prove that the neutron shell gap is reduced below Z=14. Indeed as the Z$\le$12 nuclei that lie in the island of inversion are deformed, states are very mixed and their spherical single particle (SP) energies cannot be determined there. We have to stand as much as possible $\textit{out of}$ the island of inversion to reveal the evolution of SP energies and probe the mechanism that is conducting the shell evolution in the N=20 region. Very recent experimental studies using transfer reactions have brought significant insight upon this. They took advantage of the fact that the N=14 and N=16 shell gaps are relatively large in the Ne isotopic chain to study the evolution of the neutron  N=20 and N=28 gaps at Z=10.

\begin{figure}
\centering
\includegraphics[width=8.2cm]{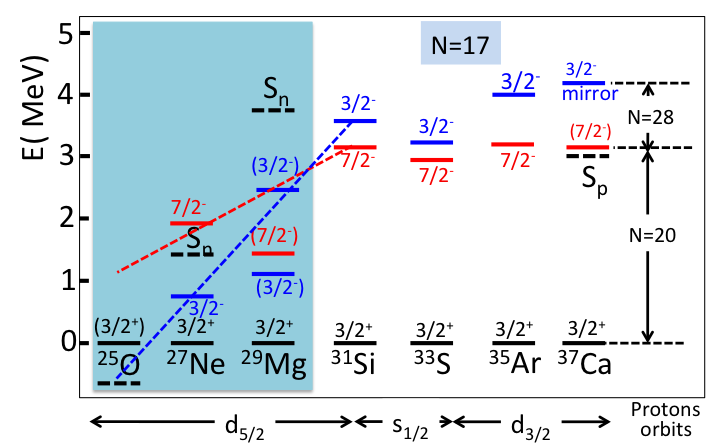}
\caption{Evolution of neutron 3/2$^+$, 7/2$^-$ and 3/2$^-$ states in the N=17 isotonic chain. The red and blue dashed lines corresponds to a simple linear extrapolations of the 7/2$^-$ and 3/2$^-$ energies towards the drip line, respectively. They show the swapping between the 7/2$^-$ and 3/2$^-$ states and the reduction of the N=20 gap but are not intended to have a significant predictive power.}
\label{N17fig}     
\end{figure}

The $^{24}$Ne and $^{26}$Ne, produced by the SPIRAL1 facility at an energy of about 10 $A$.MeV, impinged into a CD$_2$ target in which the (d,p) reaction took place to populate the 3/2$^+$, 7/2$^-$ and 3/2$^-$ states that arise mainly from neutron d$_{3/2}$, f$_{7/2}$ and p$_{3/2}$ configurations, respectively~\cite{Catf,Brown}. Protons were detected by the TIARA detector array, $\gamma$-rays by the EXOGAM detector, while the transfer-like products were identified at the focal plane of the VAMOS spectrometer. The energy, spin and spectroscopic factors (SF) of the 3/2$^+$, 7/2$^-$ and 3/2$^-$ states were obtained from the proton energy, angular distribution and cross section, respectively. SF values are in most of the cases larger than 0.6, suggesting that these states carry a significant fraction of single particle strength. It is however important to bear in mind that the forthcoming discussion  is semi-quantitative only and that detailed theoretical calculations should confirm the conclusions drawn here. Fig.~\ref{N17fig} shows the evolution of the 3/2$^+$, 7/2$^-$ and 3/2$^-$ states in the N=17 isotones based on the result of~\cite{Brown} for $^{27}$Ne. Two different regimes can be distinguished in this figure, with a turning point at Z=14, as already found for the N=20 isotones. 
Above Z=14,  it is seen that the spacing between the 3/2$^+$ and 7/2$^-$ states forming the N=20 gap is almost constant after the addition of 6 protons in the s$_{1/2}$ and d$_{3/2}$ orbits. This means that nuclear forces between the protons in these orbits and neutrons in the two orbits forming the N=20 gap  are almost the same. The spacing between the 7/2$^-$ and 3/2$^-$ states, which traditionally forms the N=28 gap (very small here $\simeq$ 1 MeV), decreases slightly between Z=14 and Z=16 and increases afterwards until Z=20. This trend is in agreement with what is predicted in the sdpf shell-model interactions~\cite{Nowa09,Gaud06}. 
Below Z=14 two major changes occur. The 7/2$^-$ and 3/2$^-$ states are swapped and the N=20 gap collapses. Even if states are not of fully single particle origin, this drastic trend should carry valuable information upon the intensity of the proton-neutron forces involved here and their spin, angular momentum and   radial overlap dependencies. To give an example on the spin dependance, while the 3/2$^+$ and 7/2$^-$ spacing is unchanged by adding four protons in the 1d$_{3/2}$ orbit from Z=16 to Z=20, it increases by about 1.2 MeV by adding four protons in the 1d$_{5/2}$ orbit from Z=10 to Z=14. An even more dramatic effect occurs for the 7/2$^-$ and 3/2$^-$ spacing which increases by 1 MeV (2 MeV) with the filling of the d$_{3/2}$ (d$_{5/2}$) orbit. Note that the results on the 3/2$^+$, 7/2$^-$ and 3/2$^-$ states in $^{25}$Ne obtained through (d,p) reaction also confirm that a reduction of the 3/2$^+$ -- 7/2$^-$ spacing and a swapping between the 7/2$^-$ and 3/2$^-$ states take place between the $^{29}$Si and $^{25}$Ne isotones~\cite{Catf}, as  between the $^{31}$Si and $^{27}$Ne isotones. We shall see in the next Section that these modifications can be intuited using known properties of the nuclear force. We shall then propose that a universal mechanism, based on the hierarchy of few nuclear forces, is expected to provide these significant shell evolutions and re-ordering around N=20, but also in other regions of the chart of nuclides.   

\section{Hierarchy of nuclear forces: toward a universal mechanism ?}
\label{hierarchy}

\begin{figure*}
\centering
\includegraphics[width=15cm]{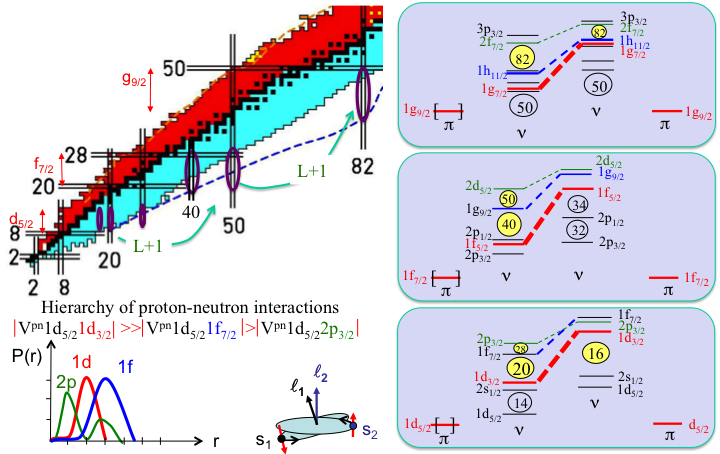}
\caption{Top Left: The regions indicated in purple ellipsoids are those discussed in the text and in the right part of the figure. Bottom left: Simplified view of the hierarchy of the proton-neutron nuclear forces based on their radial wave functions and spin dependences. Right: Schematic evolution of the shells as a function of the filling of the proton (from the bottom to top panels) 1d$_{5/2}$,  1f$_{7/2}$ and 1g$_{9/2}$ orbits. The thickness of the dashed lines connecting the neutron levels scales with the intensity of the proton-neutron interaction into play. For instance the strong proton-neutron d$_{3/2}$- d$_{5/2}$ interaction that is changing the size of the N=20 gap is symbolized by a thick red dashed-line in the bottom panel. }
\label{hierarchyfig}       
\end{figure*}

As shown above, a significant change in the energy of the neutron orbits  1d$_{3/2}$, 1f$_{7/2}$ and 2p$_{3/2}$ occurs when changing the occupancy of the proton 1d$_{5/2}$ orbit. By using simple properties of the nuclear force, we can show that the corresponding two-body proton-neutron forces V$^{pn}$ at play here follow this hierarchy $|$V$^{pn}_{1d_{5/2}1d_{3/2}} | > |$V$^{pn}_{1d_{5/2}1f_{7/2}} | > | $V$^{pn}_{1d_{5/2}2p_{3/2}}|$. Indeed the nucleon wave functions are characterized by the number of nodes $n$ of the radial part, the angular momentum $\ell$ and spin orientation with respect to the angular momentum. When the proton and neutron wave functions have the same $n$ and $\ell$ values and opposite spin orientations with respect to $\ell$, the amplitude of the corresponding interaction is maximized~\cite{Otsu}. As $|$V$^{pn}_{1d_{5/2}1d_{3/2}}$ $|$ contains all these properties, it has the largest intensity. On the other extreme, when $n$ differs, the protons and neutrons radial wave functions oscillate out of phase, hereby reducing their overlap considerably (see the schematic view of Fig.~\ref{hierarchyfig}). Added to the fact that the proton and neutron angular momenta differ by one unit and that the proton and neutron spin orientations are aligned, the $|$ V$^{pn}_{1d_{5/2}2p_{3/2}}$$|$ interaction has all reasons to be very weak. The remaining interaction $|$V$^{pn}_{1d_{5/2}1f_{7/2}}$$|$ falls in the middle, $n$ is similar for protons and neutrons, but their $\ell$ values differ by one unit, and their spin are aligned with respect to $\ell$. We let the reader find that such a clean hierarchy cannot be obtained when filling the proton 1d$_{3/2}$ orbital during which no such drastic changes in the spacing between neutron orbits is observed. This hierarchy of proton-neutron forces derived at N=17 applies as well to the regions south to $^{34}$Si and $^{42}$Si, as shown in Fig.~\ref{hierarchyfig}. Below $^{34}$Si, this accounts for the reduction of the N=20 gap and a likely swapping between the neutron 7/2$^-$ and 3/2$^-$ states. Below the deformed $^{42}$Si nucleus, the size of the N=28 gap between the f$_{7/2}$ and p$_{3/2}$ orbits is expected to be further reduced. It follows that N=28 isotones, as $^{40}$Mg, will be also deformed as seems to be indicated by the recent work on the neutron-rich $^{34-38}$Mg isotopes \cite{Door13}. Then no sharp drop in S$_{2n}$ is expected at (Z$<$14,N=28), but rather a smooth variation, bringing the drip-line further from stability as observed in Ref.~\cite{Baum}. As there is no N=20 effect as well in the Mg isotopic chain, it follows as suggested in Ref.\cite{Caur} that a merging of the two islands of inversions N=20 and N=28 is occurring there. 
Interesting extrapolations of this hierarchy of forces to other regions of the chart of nuclides can be seen in Fig.~\ref{hierarchyfig}. By  changing all involved proton and neutron angular momenta by one unit, other shell gaps are concerned. Indeed, by applying $\ell \rightarrow \ell+1$, the proton orbit becomes 1f$_{7/2}$, while the neutron orbits are 1f$_{5/2}$, 1g$_{9/2}$ and 2d$_{5/2}$. Between these neutron orbits the N=40 and 50 are formed, the intensity of which depends on the action of proton-neutron forces involved. Starting from the $^{68}$Ni nucleus, the size of the N=40 gap between 1f$_{5/2}$ (or 2p$_{1/2}$) and 1g$_{9/2}$ as well as of the N=50 gap between the 1g$_{9/2}$ and 2d$_{5/2}$ are expected to be reduced significantly when moving toward $^{60}$Ca (see for instance the predictions of Ref.\cite{Lenz13}). These latter orbits can eventually be inverted there, as proposed in the theoretical calculations of~\cite{Hagen1} that use realistic interactions. This hierarchy of forces will also impact the structure of nuclei south to $^{78}$Ni by  reducing the N=50 gap there. 

The last extrapolation ($\ell \rightarrow \ell+1$) will bring us into the region below $^{132}_{50}$Sn. This is a region in which the r-process spend enough time to build the A$\simeq$130 peak in the abundance curve of the elements. The location, height and shape of this  peak could be traced back from the neutron separation energies (S$_n$), the half-lives (T$_{1/2}$), the neutron delayed emission probabilities ($P_n$), and the neutron-capture cross sections
($\sigma_n$) of the nuclei located around the $N=82$ shell closure.  They vary with the structural evolution of the
nuclei.  Using the hierarchy $|$V$^{pn}_{1g_{9/2}1g_{7/2}}|> |$V$^{pn}_{1g_{9/2}1h_{11/2}}| >|$ V$^{pn}_{1g_{9/2}2f_{7/2}}|$ both the size of the N=82 gap and the energy between the neutron 1h$_{11/2}$ and 1g$_{7/2}$ orbits are expected to be reduced far from stability (see Fig.~\ref{hierarchyfig}). We can anticipate two important consequences for the r-process. First the reduction of the N=82 gap is expected to change the location of the r-process path, changing the shape of the r-process peak accordingly. Instead  of being all accumulated at N=82, isotopes with lower neutron number will become r-process progenitors, filling the low-mass wing of the A=130 peak. Second the $\beta$-decay half-lives T$_{1/2}$, that mainly depend on the  GT transition  $\nu$1g$_{7/2} \rightarrow \pi$1g$_{9/2}$, will be drastically shortened far from stability~\cite{Cuen,Dill}, speeding up the r-process accordingly. Indeed as T$_{1/2}$ scales with the phase factor value 1/(Q$_\beta$-E*)$^5$, where E* is the energy to which the GT transition  occurs, the reduction of the neutron 1g$_{7/2}$ -- 1h$_{11/2}$ spacing is decreasing E* and shortening the half-lives  T$_{1/2}$ of the nuclei that are paving the r-process.  Therefore  the proton-neutron interaction strongly influence the half-lives values in the N=82 isotones. 
%In very hot and neutron-dense environments, the r-process develops through neutron captures on very short timescales until reaching nuclei with sufficiently small binding energy (S$_n (A+1) \simeq $ 2-3~MeV). Typically this occurs at the major shell closures, such as the N=82 one, where a sudden drop in S$_n$ is present at A=83. There, the rate of captures is balanced by that of photodisintegration induced by the ambient photon bath of the exploding star. The process is stalled at these so-called waiting-point nuclei in a (n,$\gamma$) - ($\gamma$,n) equilibrium until $\beta$-decays occur, depleting the nucleosynthesis to the upper Z isotopic chain where subsequent neutron captures could occur. After successive $\beta$-decays and neutron captures at the N=82 closed shell (say from Z$\simeq $ 42 to 50), the process is progressively driven closer to stability where $\beta$-decay lifetimes (T$_{1/2}$) become longer. There, around the Sn isotopic chain, neutron captures are expected to be shorter than photo-disintegration and $\beta$-decay rates, driving potentially the r-process nucleosynthesis towards the next shell closure. At the end of the r-process, when the explosive conditions are over, these radioactive progenitors decay back to stability via $\beta$ or $\beta$-delayed neutronemission towards the valley of stability, forming a peak at A$\simeq$130 there. 

To finish this Section few remarks can be put forward. \emph{First}, we proposed that a hierarchy of nuclear forces is applicable to several regions of the chart of nuclides, leading to significant shell evolutions. Very recent calculations using realistic forces have been made for the O  (Ca) isotopic chain, showing that the inversion between the neutron fp (gd) shell has occurred when reaching the drip line. This simplified vision of the hierarchy of nuclear forces somehow finds echo with the properties of realistic forces.  \emph{Second}, the same forces play a role while changing  occupancies of the proton orbits 1d$_{5/2}$, 1f$_{7/2}$ and 1g$_{9/2}$ by up to 6, 8 and 10 protons, respectively. A relatively good predictive power of the evolution of shells is therefore expected far from stability, provided that the aforementioned interactions are well constrained in a given region of mass. \emph{Third}, correlations have to be taken into account to fully model the nuclei along the regions shown in Fig.~\ref{hierarchyfig}. \emph{Fourth}, in the N=82 region other shells are present between the major ones mentioned above. While they are not expected to change this global picture, they act to dilute the action of the aforementioned nuclear forces as other intermediate shells are filled. It is therefore expected that structural evolutions will be occurring more slowly in heavy nuclei. This could explain why recent isomer studies on the N=82 isotones of $^{130}$Cd (Z=48) and $^{128}$Pd (Z=46) \cite{Jung07,Wata13}  do not indicate that a change in nuclear structure has yet occurred there. Important is however to note that these two experiments are rather sensitive to proton configurations (g$_{9/2})^2$ and rather indirectly to neutron configurations. Therefore, a more direct evolution of neutron single-particle energies should be probed there in the future in order to see if the N=82 gap is reduced below Z=50. \emph{Fifth}, the same hierarchy seems also applicable to lighter nuclei to account for the evolution of the neutron 2s$_{1/2}$, 1p$_{1/2}$ and 1d$_{5/2}$ orbits when the proton 1p$_{3/2}$ is filled in the N=7 isotones between the recently studied $^{9}$He \cite{Alka} and $^{13}$C. 

\section{Nuclear forces at the drip line}
\label{drip}

In the previous Sections, it was assumed that the proton-neutron nuclear forces do not change when approaching the neutron drip-line. There, a large asymmetry is present between the proton and neutron binding energies. While the proton is deeply bound and its wave function well confined inside the nucleus, the neutron is weakly bound and its wave function can extend out of the mean nuclear radius. In such a case, the overlap of their wave functions is reduced and the effective proton-neutron interaction is weakened. This change in effective interaction depends on the binding energy of the neutron and its angular momentum. Indeed when the neutron centrifugal barrier is high as compared to the binding energy of the unbound neutron, it can be considered as quasi-bound and the effective interaction can be similar to that in the valley of stability. Several experimental works are being carried out worldwide (e.g.~\cite{Bau}) to probe the structure of unbound states and theoretical works are treating the interaction with the continuum explicitly~\cite{For}.  

\begin{figure}
\centering
\includegraphics[width=8.2cm]{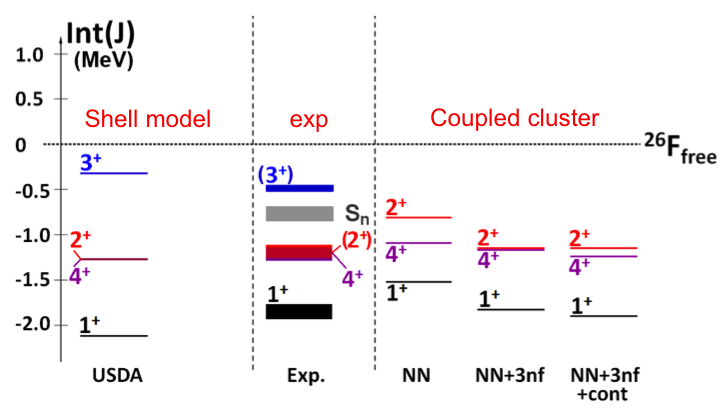}
\caption{Experimental interaction energies Int(J) of the J=1-4 states (center part) are compared to shell model calculations using the USDA interaction (left part) and to coupled cluster calculations (right part) that implement three body interactions (3nf) and the coupling to continuum (cont).}
\label{int}       
\end{figure}

As described in Ref.\cite{Lepailleur} the $^{26}_{9}$F$_{17}$ nucleus is a suitable choice for studying the proton-neutron d$_{5/2}$-d$_{3/2}$ interaction that has a decisive role for changing the N=20 gap when approaching the drip lines. The binding energies of all states of the multiplet $J=1-4^+$ arising from the $\pi$d$_{5/2} \otimes \nu$d$_{3/2}$ coupling 
in $^{26}_{9}$F$_{17}$ were determined using four experimental techniques (time of flight measurement for the J=1 ground state ~\cite{Jura07}), in-beam $\gamma$-ray spectroscopy for the J=2 state~\cite{Stan12}, M3 isomer decay study for the J=4 state~\cite{Lepailleur}, and in-beam neutron spectroscopy for the unbound J=3 state~\cite{Fran11}. The corresponding proton-neutron interaction energy, Int(J), was determined in \cite{Lepailleur}  from the difference between the experimental  binding energy of a state J in $^{26}$F and  that of the $^{24}$O+1p+1n system, in which proton and neutron are free of interaction (Int=0).  Fig. \ref{int} compares the experimental results to shell model (SM) calculations using the USDA interaction fitted to nuclei close to the valley of stability and to coupled cluster (CC) calculations that use interactions from chiral effective field theory, the treatment of three-body forces and the interaction with the continuum. Comparison with SM calculations shows that the experimental binding energy is globally weaker than expected by about 17\%, and that the multiplet of J=1-4 states is compressed by about 25\%. These discrepancies possibly point out the deficiencies of the SM that is treating nucleon interactions using a Harmonic oscillator potential basis that is not suited for taking the proton-to-neutron binding energy asymmetry and continuum effects into account. On the other hand CC calculations, in which the weak binding energy of the neutron is considered explicitly, agree with experimental results. Further studies on other nuclei should be carried out to confirm the present findings upon the proton-neutron interaction when approaching the drip-line.\\

\section{Summary and Perspectives}
\label{conc}

The field of nuclear structure far from stability is experiencing tremendous breakthroughs. Benchmarking nuclei as $^{24}$O, $^{42}$Si, $^{54}$Ca, $^{64}$Cr, $^{100}$Sn and $^{132}$Sn - to quote a few - have been or are on the way to be studied. Extremes of isospin, ranging from the proton to neutron drip lines were probed, as for instance between $^{14}_{9}$F$_{5}$ and $^{31}_{9}$F$_{22}$. When sorting out the data available to date, amazingly similar pictures arise throughout the chart of nuclides, which have been recently identified to take root from specific properties of the nuclear force and from nuclear correlations. The magic nuclei N= 8, 20 and 40 (leading to so-called HO shell gaps) all disappear far from stability,  leading to three islands of inversion through the same mechanism. It was proposed here that the reduction of the N=8, 20, and 40 shell gaps and the swapping between certain levels arise from a simple hierarchy between proton-neutron forces. This hierarchy was applied to yet unexplored regions of the chart of nuclides such as below $^{42}$Si, $^{64}$Cr, $^{78}$Ni or below $^{132}$Sn where the r-process nucleosynthesis is accumulating peaks in the r abundance curve of the elements. As for the the growing of the so-called SO shell gaps at N=14, 28 and 50 by about 2.7 MeV, it is likely due to three-body neutron-neutron forces, which were only recently implemented in theoretical models. These remarkable successes and similarities in different regions of the chart of nuclides however reach limit when looking at the evolution of SO shell gaps far from stability. For so far unknown reasons, the valence mirror nuclei built on proton and neutron SO numbers such as $^{20}_{6}$C$_{14}$, $^{42}_{14}$Si$_{28}$ and $^{132}_{50}$Sn$_{82}$, behave totally differently. The first two are deformed, while the latter is a doubly magic nucleus. It will be therefore very important to study the $^{78}_{28}$Ni$_{50}$ nucleus, which lies in between, to better understand why similar forces do not reach the same effect there.


\begin{thebibliography}{0}
\expandafter\ifx\csname natexlab\endcsname\relax\def\natexlab#1{#1}\fi
\expandafter\ifx\csname bibnamefont\endcsname\relax
  \def\bibnamefont#1{#1}\fi
\expandafter\ifx\csname bibfnamefont\endcsname\relax
  \def\bibfnamefont#1{#1}\fi
\expandafter\ifx\csname citenamefont\endcsname\relax
  \def\citenamefont#1{#1}\fi
\expandafter\ifx\csname url\endcsname\relax
  \def\url#1{\texttt{#1}}\fi
\expandafter\ifx\csname urlprefix\endcsname\relax\def\urlprefix{URL }\fi
\providecommand{\bibinfo}[2]{#2}
\providecommand{\eprint}[2][]{\url{#2}}

\end{thebibliography}


\begin{thebibliography}{}

\bibitem{Elsa34} W. Elsasser, J. de Phys. et Rad. \textbf{5} 625 (1934)  
\bibitem{Goep49} M.~Goeppert-Mayer, Phys. Rev. \textbf{75} 1969 (1949)  
\bibitem{Haxe49} O. Haxel, J.H.D. Jensen and H.E. Suess, Phys. Rev. \textbf{75} 1766 (1949)
\bibitem{Eliot54} J. P. Elliott and A. M. Lane, Phys. Rev. \textbf{96} 1160 (1954) 
\bibitem{Sorlin} O. Sorlin and M.-G. Porquet, Phys. Scr. T \textbf{152} 014003 (2013) 
\bibitem{Holt}  J. D Holt et al.,  J. Phys. G \textbf{39}, 085111 (2012)
\bibitem{Sieja} K. Sieja and F. Nowacki, Phys. Rev. C \textbf{85} 051301 (R) (2012) 
\bibitem{Talm60} I. Talmi and I. Unna, Phys. Rev. Lett. \textbf{4} 469 (1960) 
\bibitem{Thib75} C. Thibault et al., Phys. Rev. C \textbf{12} 644 (1975) 
\bibitem{Hube78} G. Hubert et al., Phys. Rev. C \textbf{18} 2342 (1978) 
\bibitem{Detr79} C. D\'etraz et al.,  Phys. Rev. C \textbf{19} 164 (1979) 
\bibitem{Guil84} D. Guillemaud-Mueller et al., Nucl. Phys. A \textbf{426} 37 (1984)
\bibitem{Camp75} X. Campi et al., Nucl. Phys. A \textbf{251} 193 (1975) 
\bibitem{Pove87} A. Poves and J. Retamosa, Phys. Lett.  B \textbf{184} 311 (1987) 
\bibitem{Warb90} W. Warburton, J. A. Becker and B. A. Brown,  Phys. Rev. C \textbf{41} 1147 (1990) 

\bibitem{N40p} M. Hannawald et al., Phys. Rev. Lett. \textbf{82}1391 (1999);  
O. Sorlin et al., Eur. Phys. J. A \textbf{16} 55 (2003)
%\bibitem{Lenzi} S.M. Lenzi, F. Nowacki, A. Poves, and K. Sieja, Phys. Rev. C \textbf{82} 054301 (2010)
\bibitem{N40}T. Baugher et al.,  Phys. Rev. C \textbf{86} 011305(R) (2012); A. Gade et al., Phys. Rev. C \textbf{81} 051304(R) (2010); P. Adrich et al., Phys. Rev. C \textbf{77} 054306 (2008); J. Ljungvall et al., Phys. Rev. C \textbf{81} 061301(R) (2010); W. Rother et al., Phys. Rev. Lett. \textbf{106} 022502 (2011); A. B\'urger et al., Phys. Lett. B \textbf{622}, 29 (2005); N. Aoi et al., Phys. Rev. Lett. \textbf{102} 012502 (2009); H. L. Crawford et al., Phys. Rev. Lett. \textbf{110} 242701 (2013)

%\bibitem{N20p}  T. Motobayashi et al., {\it Phys. Lett.} B346 (1995) 9; H. Iwasacki et al., {\it Phys. Lett.} B522 (2001) 9; B. V. Pritychenko et al., {\it Phys. Rev.} C63 (2000) 011305(R); J.A. Church et al., {\it Phys. Rev.} C72 (2005) 054320; Y. Yanagisawa et al., {\it Phys. Lett.} B566 (2003) 84; Neyens...
\bibitem{Oza00} A. Ozawa et al., Phys.~Rev.~Lett.~{\bf 84} 5493 (2000)
\bibitem{Hoff} C. R. Hoffmann et al., Phys. Lett. B \textbf{672} 17 (2009)
\bibitem{Tshoo} K. Tshoo  et al., Phys.~Rev.~Lett.~{\bf 109} 022501 (2012)
\bibitem{Huck85} A. Huck et al., Phys. Rev. C ~{\bf 31} 2226 (1985)
\bibitem{Wien} F. Wienholtz et al., Nature \textbf{498} 346 (2013)
\bibitem{Wimm} K. Wimmer et al., Phys.~Rev.~Lett.~{\bf 105} 252501 (2010)
\bibitem{Rota} F. Rotaru et al.  Phys.~Rev.~Lett.~{\bf 109} 092503 (2012)
\bibitem{Catf} W.N. Catford et al., Phys. Rev. Lett. ~{\bf 104} 192501 (2010)
\bibitem{Brown} S. M. Brown et al. Phys. Rev. C ~{\bf 85} 011302 (R) (2012)
\bibitem{Nowa09} F. Nowacki and A. Poves, Phys. Rev. C \textbf{79} 014310 (2009)
\bibitem{Gaud06} L. Gaudefroy et al., Phys. Rev. Lett. \textbf{97} 092501 (2006) 
\bibitem{Otsu} T. Otsuka, Phys. Scr. T \textbf{152} 014007 (2013) 
\bibitem{Door13} P. Doornenbal et al., Phys. Rev. Lett. \textbf{111} 212502 (2013)
\bibitem{Baum} T. Baumann et al., Nature \textbf{449} 1022 (2007)

\bibitem{Caur} E. Caurier, F. Nowacki and A. Poves,  arXiv:1309.6955v1 (2013)
\bibitem{Lenz13} S. M. Lenzi, F. Nowacki, A. Poves, and K. Sieja, Phys. Rev. C 82, 054301 (2010) 
%\bibitem{Roth} R.Roth et al., Phys. Rev. Lett. \textbf{110}  242501 (2013)

\bibitem{Hagen1} G.~Hagen et al., Phys.~Rev.~Lett. \textbf{109} 032502 (2012)

\bibitem{Cuen} J.J. Cuenca-Garcia et al. ,  Eur. Phys. J. A \textbf{34} 99 (2007)
\bibitem{Dill} I. Dillmann et al., Phys.~Rev.~Lett. \textbf{91} 162503 (2003)
\bibitem{Jung07} A. Jungclaus et al., Phys.~Rev.~Lett. \textbf{99} 132501 (2007)
\bibitem{Wata13} H. Watanabe et al., Phys.~Rev.~Lett. \textbf{111} 152501 (2013)
\bibitem{Alka} T. Al Kalanee et al., Phys.~Rev.~C {\bf 88} 034301 (2013)
\bibitem{Lepailleur} A. Lepailleur et al., Phys. Rev. Lett. {\bf 110} 082502 (2013) 
\bibitem{Bau} T. Baumann, A. Spyrou and M. Thoennessen  Rep. Prog. Phys.  \textbf{75} 036301 (2012)
\bibitem{For} C.Forss\' en et al., Phys. Scr. T \textbf{152} 014022 (2013) 


%\bibitem{Hoff08} C. R. Hoffmann  et al., Phys. Rev. Lett. {\bf 100}, 152502 (2008)
\bibitem{Jura07} B.~Jurado  et al.,  Phys.~Lett.~B. {\bf 649} 43  (2007)
\bibitem{Stan12} M. Stanoiu et al., Phys.~Rev.~C {\bf 85} 017303 (2012)
\bibitem{Fran11} N.~Frank et al., Phys.~Rev.~C {\bf 84} 037302 (2011)



%\bibitem{Hagen2} G.~Hagen et al., Phys. Rev. Lett. {\bf 108}, 242501 (2012).

\end{thebibliography}
\end{document}